\documentstyle[12pt,fleqn]{article}

\begin{document}
\thispagestyle{empty}

\vspace*{10mm}
{ \flushleft
{\large   Thermal Fluctuations of a Liquid Helium
\\Inside a Spherical Volume}\\
\vspace*{5mm}
A.V.Kirilyuk$^a$ and A.V.Zvelindovsky$^b$\\
\vspace*{5mm}
$^a$ Department of Theoretical Physics, Odessa I.I.Mechnikov
State University\\
Dvoryanskaya 2, 270026 Odessa, Ukraine\\
$^b$ Department of Biophysical Chemistry, University of
Groningen\\
Nijenborgh 4, 9747 AG Groningen, The Netherlands}

\vspace*{9mm}
 The theory of the thermal hydrodynamic 
fluctuations in bounded systems has been developed only for
non-superfluid liquids on the base of the linearized Navier-Stokes
equations [1]. However, the study of effect of confined geometry 
on correlation functions of superfluid liquid is 
important for explanation of experiments in which capillaries, pores, dense
suspensions and other objects are probed by radiation [2]. 

In this paper we present the correlation theory of the
equilibrium thermal hydrodynamic fluctuations of compressible
superfluid helium confined by the spherical cavity of radius $R$. 
The hydrodynamic fluctuations in the superfluid helium are described 
on the base of linearized two-liquids Khalatnikov's equations [3] with 
spontaneous heat fluxes and strains. For mass flow $\delta \vec{J}$ and
relative velocity $\delta \vec{w}=\vec{v}_n   - \vec{v}_s$
of normal-fluid and superfluid components these equations have form
$$
\frac{\partial \delta\rho}{\partial t}+ \nabla\cdot\delta
\vec{J}= 0, 
\ \ \rho \frac{\partial \delta s}{\partial t} +\rho_s s \nabla
\cdot \delta\vec{w}- \frac{\kappa}{T}\Delta\delta T= -\frac{1}{T}
\nabla\cdot\vec{q}, \hphantom{aaaaaaaaaaaaaaaaaaaaaaaaaaaaaaaaaaaa}
$$
$$
\frac{\partial \delta \vec{J}}{\partial t}+ \nabla\delta
p-\frac{\eta}{\rho} \Delta \delta \vec{J}-\frac{1}{\rho} (\frac13 \eta+\zeta_2) 
\nabla(\nabla\cdot\delta\vec{J})-\frac{\rho_s\eta}{\rho}\Delta\delta\vec{w}+
\frac{\rho_s}{\rho}(\rho\zeta_1 -
\frac13\eta-\zeta_2)\nabla(\nabla\cdot \delta\vec{w})=-\vec{f},
\hphantom{aaaaaaaaa}
$$
$$
\frac{\partial \delta \vec{J}}{\partial t}-\rho_n\frac{\partial
\delta \vec{w}}{\partial t}+ \nabla\delta p-\rho s\nabla\delta
T- \zeta_1\nabla(\nabla\cdot\delta\vec{J})+\frac{\rho_s}{\rho}
(-\rho\zeta_1+\rho^2\zeta_3)\nabla(\nabla\cdot\delta\vec{w})=
-\rho\nabla H, \hphantom{aaaa}
 \eqno{(1)}
$$
where
$$
\delta p=\left(\frac{\partial p}{\partial \rho} \right)_T
\delta\rho +\left(\frac{\partial p}{\partial T}
\right)_{\rho}\delta T,  \ \ \delta s= 
\left(\frac{\partial s}{\partial \rho} \right)_T \delta \rho
+\left(\frac{\partial s}{\partial T}\right)_{\rho}\delta T.
\hphantom{aaaaaaaaaaaaaaaaaaaaaaa}
$$
Here $s$ is the entropy per unit mass, $p$ is the pressure;
$\kappa$, $\eta$, and $\zeta_1, \zeta_2, \zeta_3$ are the
heat conductivity, the "first" viscosity, and the "second"
viscosity coefficients, respectively; $ \rho_s, \rho_n,$ and $
\rho$ are the superfluid,  normal-fluid, and total densities.
The system (1) is written with account of the Onsager symmetry principle
for the kinetic coefficients. All kinetic coefficients and
thermodynamic derivatives in (1) assumed to be independent of
coordinates. By (1) the fluctuation fields of mass flow, mass density
and other variables are linear functionals of random forces in 
the right parts of these equations. 

 Assuming that the rigid-cavity surface is motionless we apply zero boundary 
conditions for the perpendicular component of mass flow $\delta \vec{J}$ and
temperature fluctuations $\delta T$ in the liquid helium, and the stick 
boundary conditions for the normal component of velocity [4]
$$
\delta J_r(\vec{r}, t)=0,\ \ \vec{v}_n(\vec{r}, t)=0,\ \ \delta
T(\vec{r}, t)=0\      \ ( r=R ). \hphantom{aaaaaaaaaaaaaaaaaaaaaaaaaaaaa} \eqno{(2)}
$$

We present the mass flow, relative velocity,
mass density and the temperature fluctuations in the form 
$$
\delta \vec{J}( \vec{r}, t)=\sum \limits_{\lambda}
\left(J_{\lambda}^L(t) \vec{L}_{\lambda}(\vec{r})+J_{\lambda}^M(t)
\vec{M}_{\lambda}(\vec{r})+J_{\lambda}^N(t)
\vec{N}_{\lambda}(\vec{r}) \right), \hphantom{aaaaaaaaaaaaaaaaaaaa} \eqno{(3)}
$$
$$
\delta \vec{w}( \vec{r}, t)=\sum \limits_{\lambda}
\left(W_{\lambda}^L(t) \vec{L}_{\lambda}(\vec{r})+W_{\lambda}^M(t)
\vec{M}_{\lambda}(\vec{r}) +W_{\lambda}^N(t)
\vec{N}_{\lambda}(\vec{r}) \right), \hphantom{aaaaaaaaaaaaaaaaa} \eqno{(4)}
$$
$$
\delta \rho( \vec{r}, t)=\sum\limits_{\lambda}
R_{\lambda}(t)j_n(kr)Y_{mn}( \theta, \varphi),
\hphantom{aaaaaaaaaaaaaaaaaaaaaaaaaaaaaaaaaaaaa} \eqno{(5)} 
$$
$$
\delta T( \vec{r}, t)=\sum\limits_{\lambda}
T_{\lambda}(t)j_n(kr)Y_{mn}( \theta, \varphi),
\hphantom{aaaaaaaaaaaaaaaaaaaaaaaaaaaaaaaaaaaaa} \eqno{(6)}
$$
where $\lambda=\{n,l,m\}$ and 
$$
\vec{L}_{mn}(\vec{r})= \frac{1}{k_{\lambda}} \nabla(Y_{mn}(
\theta,\varphi)j_n(kr)), \hphantom{aaaaaaaaaaaaaaaaaaaaaaaaaaaaaaaaaaaaaaaaaaaa} 
$$
$$
\vec{M}_{mn}(\vec{r})=rot( \vec{r}\,Y_{mn}(\theta,\varphi)
j_n(kr)), \ \ \vec{N}_{mn}(\vec{r})= \frac{1}{k_{\lambda}}
rot\,\vec{M}_{mn}(\vec{r}). \hphantom{aaaaaaaaaaaaaaaaaaa} \eqno{(7)}
$$
Here $Y_{mn}(\theta,\varphi)$ are the spherical functions and 
$j_n(k_{\lambda} r)$ are the spherical Bessel functions [5]. Note that 
$\vec{L}_{mn}(\vec{r}),\ \ \vec{M}_{mn}(\vec{r}),\
\ \vec{N}_{mn}(\vec{r})$ form the system of pairwise orthogonal functions.  

From the boundary conditions (2) we find out eigenvalues $k_{\lambda}$: 
$$
j_n( \beta_{nl})=0,\ \ k_{ \lambda}= \beta_{nl}/R,
\hphantom{aaaaaaaaaaaaaaaaaaaaaaaaaaaaaaaaaaaaaaaaaaaaaa} \eqno{(8)} 
$$
$$
\gamma_{nl}j_n^{ \prime}( \gamma_{nl})+ \alpha j_n(
\gamma_{nl})=0,\ \ \alpha =-n,n+1,\ \ k_{\lambda}= \gamma_{nl}/R.
\hphantom{aaaaaaaaaaaaaaaaaaaaaa} \eqno{(9)} 
$$

We present also the longitudinal and transverse parts of the random viscous
force $ \vec{f}$ and heat fluxes $\vec{q}$  in the form 
of expansion (3) with coefficients $f_{ \lambda}^a$ and $Q_{ \lambda}^a$, 
respectively, $a=M, N, L$, and the random
potential $H$ - as in eq.(5) with coefficients $H_{ \lambda}$.

It is convenient to use the Fourier representation for random 
$\delta \vec{J}$, $\delta \vec{w}$, $\delta \rho$, $\delta T$ 

$$
x(t)= \int\limits_{-\infty}^{\infty}x_{ \omega}exp[-i \omega t]d \omega.
\hphantom{aaaaaaaaaaaaaaaaaaaaaaaaaaaaaaaaaaaaaaaaaaaaaa} \eqno{(10)} 
$$
Since the equations for the
Fourier-amplitudes of the mass flow $J_{ \lambda \omega}^b$ and
relative velocity $W_{ \lambda \omega}^b$ fluctuations $(b=M, N)$
do not contain components of other fields we immediately obtain that
$$
J_{ \lambda \omega}^{M,N}= \rho_n W_{ \lambda \omega}^{M,N}=
\frac{f_{ \lambda \omega}^{M,N}}{i \omega -k_{ \lambda}^2 \eta /
\rho_n} \   \ . \hphantom{aaaaaaaaaaaaaaaaaaaaaaaaaaaaaaaaaaaaaaaa} \eqno{(11)}
$$ 
Ignoring the products of dissipation coefficients, fluctuation forces  
and keeping only the terms linear in the dissipation coefficients  we get

$$
D_{ \lambda}J_{ \lambda \omega}^L=-\frac{1}{T} \omega^2 k_{ \lambda}^2
\rho_n \left(\frac{\partial p}{\partial T} \right)_{\rho} Q_{
\lambda \omega}^L -i \omega k_{\lambda}^3 \rho_s \rho s
\left(\frac{\partial p}{\partial T} 
\right)_{\rho} H_{ \lambda \omega}+\hphantom{aaaaaaaaaaaaaaaaaaaaaaaa}
$$
$$
+i \omega \left(k_{\lambda}^2
\rho_s s \left(\frac{\partial p}{\partial T}
\right)_{\rho}-k_{\lambda}^2 \rho_s \rho s^2+ \omega^2 \rho_n
\rho \left(\frac{\partial s}{\partial T}\right)_{\rho}
\right)f_{ \lambda \omega}^L, \hphantom{aaaaaaaaaaaaaaaaaaaaaaaa}
\eqno{(12)}  
$$ 
where 
$$
D_{ \lambda}=- \rho_n \rho \left(\frac{\partial s}{\partial
T}\right)_{\rho} \, \prod\limits_{j=1}^2 ( \omega -k_{ \lambda}u_j+i \alpha_{j \lambda}
u_j)( \omega +k_{ \lambda}u_j+i \alpha_{j \lambda} u_j) \hphantom{aaaaaaaaaaaaaaaaaaa} \eqno{(13)}
$$
and $u_1$, $u_2$ are the first- and second-sound velocities
for the infinite space [3]; $ \alpha_{1
\lambda}$, $ \alpha_{2 \lambda}$ are the attenuation
coefficients of the first- and second-sound, respectively.
Ignoring terms proportional to $ \rho_s
\left(\frac{\partial \rho}{\partial T}\right)_p$ and
$\frac{\kappa}{c_p} \left(\frac{c_p}{c_v}-1 \right)
\frac{u_2^2}{u_1^2}$ 
for the first- and second-sound attenuation
coefficients we obtain 
$$
\alpha_{1 \lambda}=\frac{k_{\lambda}^2}{2 \rho u_1}
\left[
\frac{4}{3} \eta+ \zeta_2 
+\frac{\kappa}{c_p} \left(\frac{c_p}{c_v}-1 \right)
\right], \hphantom{aaaaaaaaaaaaaaaaaaaaaaaaaaaaaaaaaaaaaaaa}
$$
$$
\alpha_{2 \lambda}=\frac{k_{\lambda}^2}{2 \rho
u_2}\frac{\rho_s}{\rho_n} \left[\frac{4}{3} \eta+ \zeta_2- 2 \rho
\zeta_1 + \rho^2 \zeta_3 +\frac{\rho_n \kappa}{\rho_s c_p}
\right], \hphantom{aaaaaaaaaaaaaaaaaaaaaaaaaaaa} \eqno{(14)} 
$$  
where $c_p$, $c_v$ are the specific heats at constant pressure
and volume, respectively. The expansion coefficients for the excess 
mass density can be expressed in terms of $J_{ \lambda}^L$.

From the expression for the rate of energy dissipation in the liquid HeII 
confined in the sphere volume $V$ 
$$
Q(t)=\int\limits_{V}dV\left(
-\vec{v}_n\vec{f}+\rho_s\delta\vec{w}\,\nabla H - \frac{1}{T}
\delta T \ \nabla\cdot\vec{q} \right)
\hphantom{aaaaaaaaaaaaaaaaaaaaaaaaaaaa} \eqno{(15)}
$$
we find that
$$
Q_{ \omega}= \sum\limits_{a, \lambda} \left(-\frac{1}{\rho}(
\rho_s W_{ \lambda \omega}^a+J_{ \lambda \omega}^a) \Lambda_{
\lambda}^a f_{ \lambda \omega}^a +\frac{1}{3}k_{ \lambda} \rho_s
W_{ \lambda \omega}^L \Lambda_{
\lambda}^L H_{ \lambda \omega} \right),\ \ a=L,M,N,
\hphantom{aaaaaaa} \eqno{(16)}  
$$
where $ \Lambda^a= \int\limits_{V}d \vec{r}\, | \vec{a} (
\vec{r})|^2$, $a=M,N,L$ are the normalization factors.

Using the Langevin equations (11), (12) we construct as in [6] the
matrix of the generalized susceptibilities. And then by
FDT find the spectral densities of equilibrium thermal fluctuations for the
expansion amplitudes of the hydrodynamic fields. This resulted in
$$
<J_{\lambda}^L J_{\lambda^{\prime}}^{L\, *}>_{\omega}=
\mbox{\rm Re }\frac{\Theta(\omega,T)\delta_{\lambda
\lambda^{\prime}}}{\Lambda_{\lambda}^L \pi}
\frac{i\omega}{D_{\lambda}}\rho^2 \rho_n 
\left(\frac{\partial s}{\partial
T}\right)_{\rho}(k_{\lambda}^2u_2^2- \omega^2), 
\hphantom{aaaaaaaaaaaaaaaaaaaaaaaa}  
$$
$$
<J_{\lambda}^b J_{\lambda^{\prime}}^{b\, *}>_{\omega}=
\mbox{\rm Re }\frac{\Theta(\omega,T)\rho\delta_{\lambda
\lambda^{\prime}}}{\Lambda_{\lambda}^b\pi(-i\omega+k_{\lambda}^2\eta/\rho_n)},
\ \  b=M, N, \hphantom{aaaaaaaaaaaaaaaaaaaaaaaaaa} \eqno{(17)}  
$$
where $\Theta(\omega,T)$ is the expression [6] for the average
energy of the quantum oscillator.

The result differs from the spectral densities for an infinite
superfluid by only the discrete behaviour of the wave numbers
$k_{\lambda}$. Factorisation of the coefficient $D_{\lambda}$ in
 (17) allows us to decompose the expressions for the pair correlation
function of $J_{\lambda}^L$ into the simple fractions. 
And the spectral densities of the
$J_{\lambda}^L$ as in the case of an infinite medium [7], consist
of two contributions
$$
<|J_{\lambda}^L|^2>_{\omega}=
\frac{2\rho\Theta(\omega,T)\omega^2 (k_{\lambda}^2
u_2^2- \omega^2)}{\pi \Lambda_{\lambda}^L k_{\lambda}^4 (u_1^2-u_2^2)^2}
\left\{
\frac{(\alpha_{1\lambda}u_1-\alpha_{2\lambda}u_2)
(\omega^2-k_{\lambda}^2 u_1^2)-\alpha_{1\lambda}u_1
k_{\lambda}^2(u_1^2-u_2^2)}{[(\omega -k_{\lambda} u_1)^2+
\alpha_{1\lambda}^2 u_1^2]
[(\omega +k_{\lambda} u_1)^2+\alpha_{1\lambda}^2 u_1^2]}
+\right.\hphantom{aaaaaaaaaaaaaaaaaaaaaa} 
$$
$$
+
\left.
\frac{(\alpha_{2\lambda}u_2-\alpha_{1\lambda}u_1)
(\omega^2-k_{\lambda}^2 u_2^2)+\alpha_{2\lambda}u_2
k_{\lambda}^2(u_1^2-u_2^2)}{[(\omega -k_{\lambda} u_2)^2+
\alpha_{2\lambda}^2 u_2^2]
[(\omega +k_{\lambda} u_2)^2+\alpha_{2\lambda}^2 u_2^2]}
\right\}, \ \ k_{\lambda}=\gamma_{nl}/R. \hphantom{aaaaaaaa} \eqno{(18)}
$$
% where $k_{\lambda}=\gamma_{nl}/R$.

Now it is easy to construct the spectral densities of thermal
fluctuations of the mass flow and the mass density hydrodynamic fields
$$
<\delta\vec{J}(\vec{r},t)\delta\vec{J}(\vec{r}\,^{\prime},t^{\prime})>_{\omega}=
\sum\limits_{\lambda, A}<|J_{\lambda}^A|^2>_{\omega}
\vec{A}_{\lambda}(\vec{r})\vec{A}_{\lambda}^{*}(\vec{r}\,^{\prime}),
\ \ A=M,L,N,
\hphantom{aaaaaaaaaaaa}\eqno{(19)}
$$
$$
<\delta\rho(\vec{r},t)\delta\rho(\vec{r}\,^{\prime},t^{\prime})>_{\omega}=
\sum\limits_{\lambda}<|R_{\lambda}|^2>_{\omega}
j_n(k_{\lambda}r)j_n(k_{\lambda}r^{\prime})Y_{mn}(\theta,\varphi)
Y_{mn}(\theta^{\prime},\varphi^{\prime}),
 \hphantom{aaaaaa}\eqno{(20)}
$$
where  
$$
<|R_{\lambda}|^2>_{\omega}=\frac{k_{\lambda}^2}{\omega^2}
<|J_{\lambda}^L|^2>_{\omega}. 
 \hphantom{aaaaaaaaaaaaaaaaaaaaaaaaaaaaaaaaaaaaaaaaaaa} \eqno{(21)}
$$

We have separated the first- and second-sound contributions to
the mass density - mass density correlation function. The correlation functions
are presented as the sum of the series in different
spherical harmonics and roots of the transcendental equations (8), (9). 
As mentioned above, the difference between this case and that of 
infinite space is only in the discrete nature of the wave
numbers $k_{\lambda}$. Each term in this series generates two
Mandelstam-Brillouin doublets, which are caused by quasiphonons
of two kinds, one having the first- and the other one the second-sound
speed. Each components of the doublets are of
Lorentz type, their widths being given by
$\alpha_{1 \lambda} u_1$ and $\alpha_{2 \lambda} u_2$, and the
spectral densities of the thermal fluctuations of mass flow and
mass density  can be presented as combinations of infinite
numbers of such Lorentzians. 
\vspace*{3mm}
{ \large \bf \flushleft References}

\vspace*{2mm}
\noindent
1. A.V. Zvelindovsky and A.V. Zatovsky, Nuovo Cimento D, 19(1997)
725. \\
2. D.L.Johnson, J.Phys.: Condens.Matter, 2(1990) SA449. \\
3. I.M.Khalatnikov, Introduction to the Theory of Superfluidity,
Benjamin, New York, 1965. \\
4. L.D.Landau and E.M.Lifshitz, Fluid Mechanics, Pergamon,
Oxford, 1982. \\
5. P.Morse and H.Feshbach, Methods of Theoretical Physics,
Vol.2, McGraw-Hill, New York, 1953. \\ 
6. S.M.Rytov, Yu.A.Kravtzov, V.I.Tatarsky, Introduction to the
Statistical Radiophysics, vol.2, Nauka, Moscow, 1978(in
Russian). \\
7. Seth J. Putterman, Superfluid Hydrodynamics, North-Holland,
Amsterdam, 1974.  

\end{document}